\documentclass{article}
\usepackage{spconfa4,amsmath,graphicx}
\usepackage[utf8]{inputenc}

\title{Deep Denoising for Hearing Aid Applications}

\twoauthors
  {M. Aubreville\sthanks{M. Aubreville is also affiliated with Sivantos GmbH, Erlangen, Germany.}, K. Ehrensperger, A. Maier\sthanks{A. Maier is the last author of this paper.}}
	{Pattern Recognition Lab, Computer Science\\Friedrich-Alexander-Universität Erlangen-Nürnberg\\Erlangen, Germany}
  {T. Rosenkranz, B. Graf , H. Puder\sthanks{H. Puder is also affiliated with TU Darmstadt, Graduate School of Computational Engineering, Darmstadt, Germany.}}{Sivantos GmbH\\
	Research and development\\
	Erlangen, Germany}
\begin{document}
\maketitle
\begin{abstract}
  Reduction of unwanted environmental noises is an important feature of today's hearing aids (HA), which is why noise reduction is nowadays included in almost every commercially available device. The majority of these algorithms, however, is restricted to the reduction of stationary noises. 
  
  In this work, we propose a denoising approach based on a three hidden layer fully connected deep learning network that aims to predict a Wiener filtering gain with an asymmetric input context, enabling real-time applications with high constraints on signal delay. The approach is employing a hearing instrument-grade filter bank and complies with typical hearing aid demands, such as low latency and on-line processing. It can further be well integrated with other algorithms in an existing HA signal processing chain.
  
  We can show on a database of real world noise signals that our algorithm is able to outperform a state of the art baseline approach, both using objective metrics and subject tests. 
\end{abstract}
\section{Introduction}
\label{sec:intro}

With societies aging globally, hearing loss is becoming a very common problem worldwide. 
The development of a hearing loss is typically accompanied by increasing difficulty to discriminate speech from noise in challenging situations. Consequently, besides amplification, modern hearing instruments deliver a great spectrum of algorithmic possibilities to enhance hearing, especially hearing of speech signals in noisy environments. 
  
The signal-to-noise ratio is typically improved in hearing instruments by making use of directional microphones \cite{Ricketts:1999wh,KamkarParsi:tp}. Directional processing, however, has the inherent side-effect, that the target speaker needs to be in a defined direction, often in the frontal hemisphere, and that sounds emitted from other angles will likely be attenuated. While, for many situations, this assumption can be safely made, it can also be disturbing in other situations, where the target source cannot be assumed to be coming from the front. Furthermore, for some instruments directional processing is not an option due to limitations in size and power consumption, e.g. devices to be inserted deeply into the ear canal.

\section{State of the Art}
Single-channel noise reduction aims to solve this problem, making use of a single microphone signal only. Most noise reduction schemes, however, use limited signal properties of noisy environments, effectively only exploiting first and second order statistics and targeting at steady state noises, resulting in difficulties when dealing with non-stationary background signals. To overcome this, codebook-based noise reduction was proposed by Kuropatwinski \cite{Kuropatwinski:jz}. This approach, however, lacked robustness -- a property very important for industry-grade medical products, and was combined with a recursive noise tracker by Rosenkranz and Puder to overcome this limitation~\cite{Rosenkranz:2012jz}. 
The use of longer context audio segments with neural networks has been proposed by Hermansky and Sharma \cite{hermansky1999temporal} and has found applications in automatic speech recognition \cite{univis90500975}. 
In recent years, deep learning methods and deep neural networks (DNNs) have become increasingly popular in speech recognition \cite{Dahl:dx} and speech enhancement. Since deep networks are able to learn a complex, nonlinear mapping function, this makes them ideal candidates for noise reduction tasks, where complex priors (speech and distortion signal properties) must be modeled. 

Auto-encoder networks have been used by multiple authors also in the field of audio denoising. Lu \textit{et al.} were amongst the first to report about successful use of denoising auto-encoders for speech recognition systems \cite{Lu:2013vr}. 
Xia \textit{et al.} used a denoising auto-encoder to calculate an estimate of clean speech which is then used in a traditional Wiener filtering approach \cite{bib:xia:WDA}. To cope with missing loss sensitivity at high frequencies, they introduce a frequency-dependent weight, which effectively adjusts the learning rate at the last layer of the autoencoder. However, as Kumar \textit{et al.} have highlighted, these studies lack realistic noise scenarios, as the same kind of noise was used for training as for testing. 

When the network is used for direct speech signal estimation, a high correlation between the loss function and human perception should be demanded. This common problem was addressed by Pascual \textit{et al.}, who proposed to use generative adversarial networks, in effect using discrimination loss instead of a minimum squared error that is typically used in other works \cite{Pascual:2017ug}. 

Besides auto-encoders, traditional fully-connected network topologies have been used for audio denoising. 
Xu \textit{et al.} have proposed a DNN-based regression approach for frequency-bin wise enhanced signal amplitude prediction in a filter bank setting, using the noisy phase for signal reconstruction \cite{Xu:2014kl}. In their work, they have also shown that the DNN can profit from the inclusion of a longer acoustic context. 

A very important property of a hearing assistance device is its overall latency, especially when it comes to mild to moderate hearing losses, where acoustic coupling is typically open and thus a strong component of the direct sound is reaching the ear drum. Here, comb filter effects are introduced by the superposition of processed signal and direct signal. Latencies of below 10\,ms are typically tolerated in this regard, with lower latencies being tolerated better by subjects. This places high demands on delay on every part of the signal processing chain, especially the filter banks and noise reduction schemes.

This, together with the constraint for online processing, was not given for any of the state of the art, which builds the motivation for this work.

\section{Material and Methods}

We used 49 real-world noise signals, including non-stationary signals, recorded at various places in Europe using hearing aid microphones in a receiver in the canal-type hearing instrument shell (Signia Pure 312, Sivantos GmbH, Erlangen, Germany) using calibrated recording equipment at a sampling rate of 24\,kHz. The signals have been mixed with German sentences (N=260) from the EUROM database \cite{bib:eurom:database}, which has been upsampled to 24\,kHz for this purpose. 

\subsection{Dataset Generation}

Since the noise conditions in our dataset have been recorded in real situations, levels should not be modified significantly. The signal mixing process can be expressed as:

\begin{equation}
	x = g_\textrm{L} \left( n_0 + g_\textrm{S} s_0 \right) = g_\textrm{L} n_0 + g_\textrm{L} g_\textrm{S}  s_0 = s + n
\end{equation}

where the original speech signal $s_0$ is adjusted in level to reach a defined SNR using $g_\textrm{S}$. 
For data augmentation, we adapt the idea of Kumar \textit{et al.}\cite{bib:kumar:SpeechEnhancementMultiNoiseDNN} and combine up to four noises with different offsets within the original files into the noise mixture $n_0$. 
 The noise mixture is adjusted in level ($g_L \in \{-6, 0, 6\}\,\mathrm{dB}$) to increase dataset variance, yielding a signal mixture with realistic level information. For training, signals with SNRs of $\{-100,5,0,5,10,20\}\,\mathrm{dB}$ were generated with an equal distribution. 

The train-validation-test split for our machine learning approach was done on original signal level, i.e. no speech or noise signal that is contained in the training set is part of the validation or test set.

\subsection{Signal Processing Toolchain}
\begin{figure}[t]
  \centering
  \includegraphics[width=\linewidth]{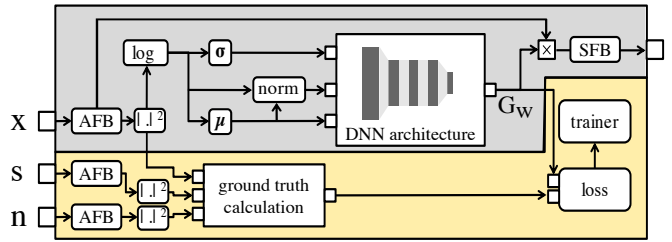}
  \caption{Processing chain during DNN training. Signals are fed into an analysis filter bank, the mixture is being normalized and statistical values are provided to the network.}
  \label{fig:processingChain}
\end{figure}

Our toolchain, as depicted in Fig. \ref{fig:processingChain}, starts with analysis filter banks (AFB) that are used to process the clean speech signal $s$, the noise signal $n$ and the noisy mixture $x$. A standard uniform polyphase filter bank with 48 frequency bins designed for hearing aid applications is utilized for this \cite{bauml2008uniform}, yielding signals $X(k,f)$, $S(k,f)$ and $N(k,f)$ with time-index $k$ and frequency index $f$. 

As a first subsequent step after the filter bank, the log power spectrum is calculated, followed by a normalization step. Then, the signal including the temporal context is fed into a fully connected network topology with 3 hidden layers and 2048 nodes per layer. Finally, a 48 channel gain vector $G_w$ is being predicted and applied to the noisy signal $X(k,f)$, which is in turn synthesized again (SFB). This structure is especially suited for hearing instrument applications, as gain application can well be combined with other algorithms working on the signal chain, such as automatic gain control for hearing loss compensation, and could be interpreted as known operator learning \cite{maier2017precision}.

\subsection{Asymmetric Temporal Context}
As previously stated, one constraint in hearing instrument signal processing is the limited latency a device is allowed to produce \cite{bauml2008uniform}. Besides the filter bank design, this constraint also strongly restricts the algorithmic lookahead for noise reduction schemes. We assumed an overall latency of $8\,$ms to be tolerable, where approx. $6\,$ms are consumed by the analysis and synthesis filter bank. Context information from past samples is in turn only limited by the memory and processing constraints of the instrument, which typically scales with available chip technology. 
The time context of this work can thus be divided into 2 components:

\begin{equation}
	\tau = \tau_1 + 1 + \tau_2
\end{equation}
where $\tau_1$ is the look back time constant, and $\tau_2$ is the lookahead time constant. Together with the current frame they define the input matrix to the network. For our setup, $\tau_1 >> \tau_2$.

\subsection{Normalization}
Normalization plays a key role in deep learning, since it positively influences convergence behavior and aids generalization. Given the filter bank representation $X(k,f)$ of the input signal $x(n)$, the signal is normalized as:

\begin{equation}
\begin{split}
	X_\mathrm{norm}(k,f) &= X(k,f) - \frac{\sum_{i=-\tau_1}^{\tau_{2}} X(k+i,f)}{\tau}\\ &= X(k,f) - \mu(f)
\end{split}
\end{equation}

This frequency bin-wise normalization is thus calculated on the available buffer, given the temporal context. This normalization scheme requires no global information and thus enables on-line processing.
The calculated mean value vector $\boldsymbol{\mu}$ is, alongside with the frequency bin-wise standard deviation $\boldsymbol{\sigma}$, enhancing the model's input as further contextual information.

\subsection{Network Details and Training}

We observed that network convergence as well as RMSE loss was positively influenced by using rectified linear unit activation functions within the hidden layers. The system was trained on a Nvidia Titan Xp GPU using TensorFlow and the Adam optimizer with an initial learning rate of $10^{-5}$ for $10$ epochs.  

\begin{figure}
	\includegraphics[width=\linewidth]{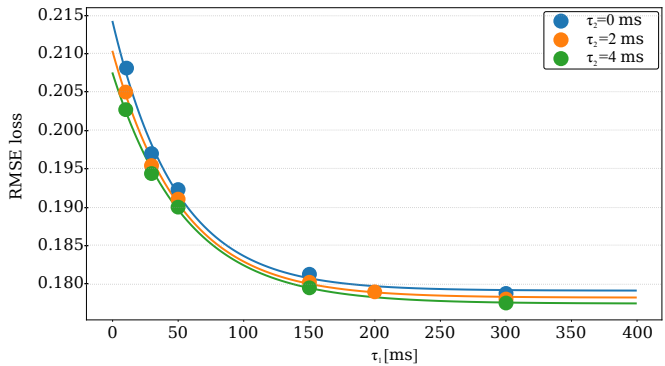}
	\caption{Validation loss in dependency of lookback time $\tau_1$ and lookahead time $\tau_2$ after 10 epochs of training. Additionally, an exponential curve fit is shown.}
	\label{context_loss}
\end{figure}

\section{Results and Discussion}

\begin{figure}
	\includegraphics[width=\linewidth]{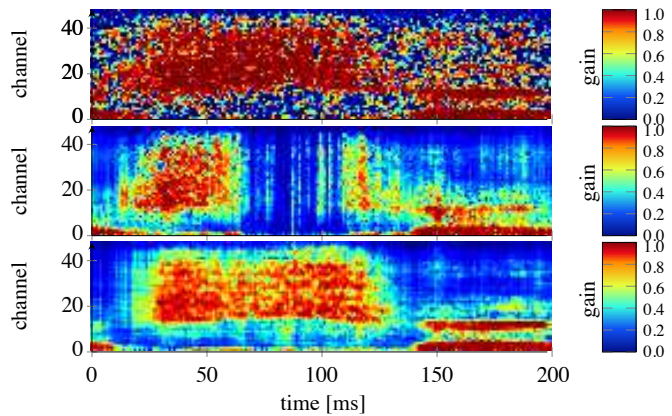}
	\caption{Exemplary comparison of a fricative with the ground truth comparing different temporal contexts. Top: ground truth, middle: $t_1=30\,\mathrm{ms}$, bottom: $\tau_1=200\,\mathrm{ms}$.}
	\label{compare_tau}
\end{figure}

When comparing the loss for different time constants of the asymmetric context (Fig. \ref{context_loss}), we find that the network profits from increased information from both past, and future. The benefit w.r.t. past context saturates at around 200\,\dots300\,ms. This is in line with the findings of Xu \textit{et al.}, and could hint at the rate of speech syllables ($\approx 4\,$Hz) \cite{houtgast1985review}. Looking at the example of a fricative (Fig. \ref{compare_tau}) further hints towards this interpretation: With almost no energy at the low frequencies, the network with greater context is still able to differentiate between the noisy fricative and a real noise signal.

We compared the results by our method with a state-of-the-art noise reduction scheme, based on recursive minimum tracking \cite{bib:hansler:fancytrack} that is being applied in commercially available hearing instruments. Another comparison was made against an idealized scheme, where the optimal Wiener gain is applied, and as anchor a badly tuned minimum statistics estimator. For all settings, the maximum attenuation of the algorithms was limited to $14\,\mathrm{dB}$.
 
\subsection{Objective Metrics}
\begin{figure}
	\includegraphics[width=\linewidth]{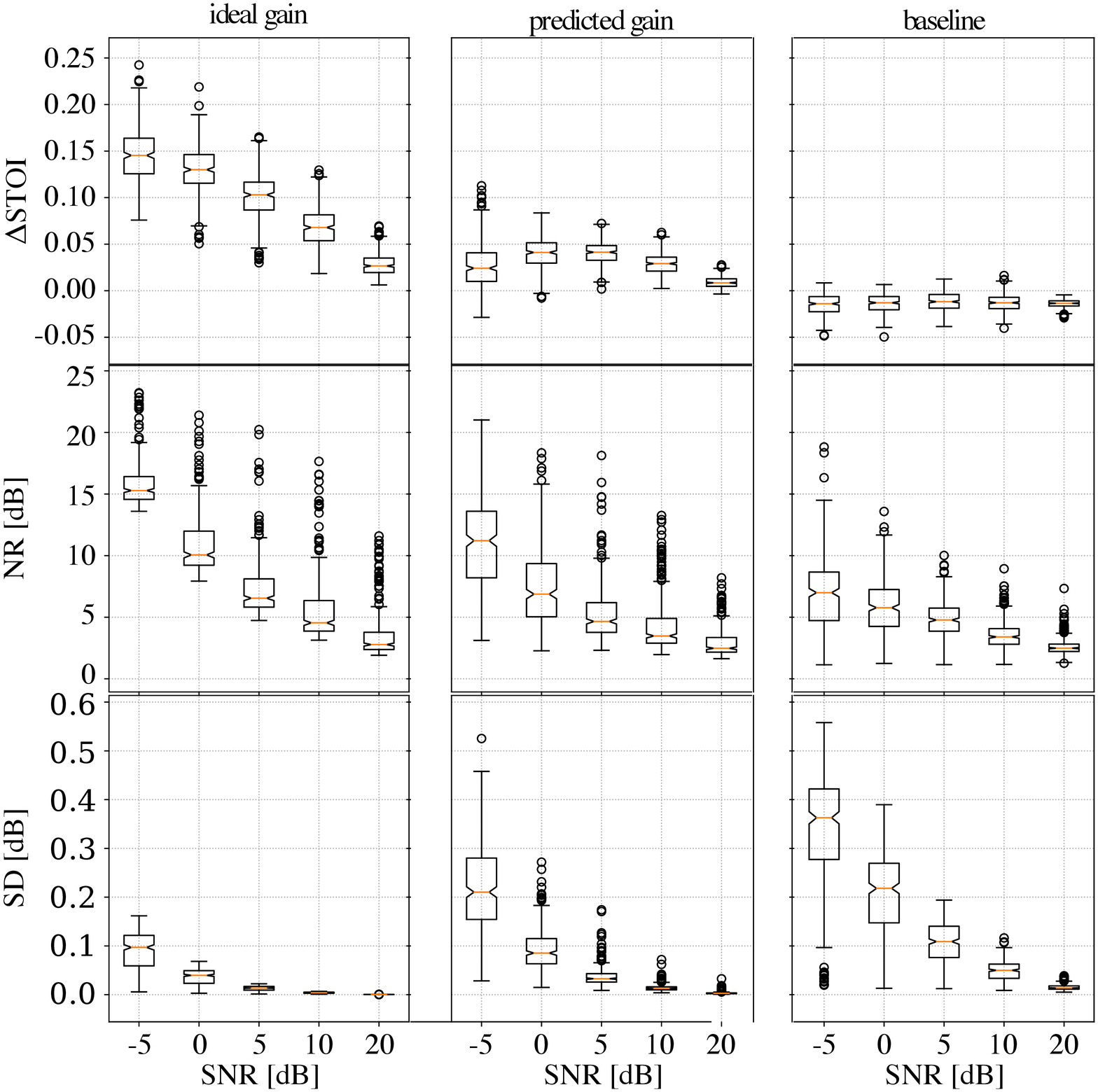}
	\caption{Objective measures of ideal Wiener gains, DNN prediction ($\tau_1=200\,\mathrm{ms},\tau_2=2\,\mathrm{ms}$) and recursive minimum tracking baseline. Top: differences in STOI index, middle: noise reduction (NR), bottom: speech distortion (SD).}
	\label{ObjectiveEvaluation}
\end{figure}

For numeric assessment of our complete test data set, we use the short-term objective intelligibility metric from Taal \textit{et al.} \cite{bib:taal:STOI}. To improve visibility of the results, we present the difference between the original noisy signal and the enhanced signal, denoted as $\Delta$STOI (see Fig. \ref{ObjectiveEvaluation}). For the DNN approach in low SNR conditions, we see that $\Delta$STOI is on average close to zero, indicating no gains in intelligibility. For SNR closer to real world conditions, the value improves with saturating effects for higher SNRs. For the baseline approach, the average value is consistently negative. 

Following the formulation of Chen \textit{et al.}, we further include the noise reduction (NR) and speech distortion (SD) metrics for our evaluation \cite{bib:chen:sdnr}. Comparing the baseline to the predicted gain using the DNN, we find that a higher noise reduction can be achieved while at the same time producing lower speech distortion.

\subsection{Subjective Evaluation}

Even though objective figures like STOI or others are designed to correlate well with perception, the validity of such measures can always be questioned for the actual noise reduction scheme that is being applied. Further, quality ratings of a noise reduction system should always be made relative to an upper boundary, limited by the SNR and the achievable optimum performance given the general approach.

In audio coding, the evaluation is subject to a similar problem, in this case that the original audio signal might be of a mediocre quality or might have signal parts that are perceived as better in the processed than the original signal given a non-optimal encoding. For this domain, the MUlti Stimulus test with Hidden Reference and Anchor (MUSHRA) \cite{liebetrau2014revision} was developed and is being widely used. 

We apply this test also for the domain of noise reduction, because we find a similar setup: We also have a potentially perceived non-optimal reference signal, we can derive an anchor signal that should be perceived as being of worse quality, and we want to compare multiple signals relative to each other. For our case, especially the comparison with the reference is of great importance, since a Wiener filter-based scheme will always have the noisy phase in the output signal and is thus restricted in quality for poor SNR conditions. 

The test was carried out at Sivantos R\&D site in Erlangen, Germany. The vast majority of all (N=20) subjects participating in the test were audio engineering professionals and thus highly qualified for signal quality assessment. Signals were presented over a web interface (webMUSHRA implementation of Schoeffler \textit{et al.}.  \cite{schoeffler2015towards}) using headphones in a calm office environment setting. 
For each SNR condition, 4 input signals of 12\,s length were randomly picked from the test dataset and processed for all test conditions. To not include initialization of the recursive minimum tracker baseline, all signals were cut after initialization.

\begin{figure}
	\includegraphics[width=\linewidth]{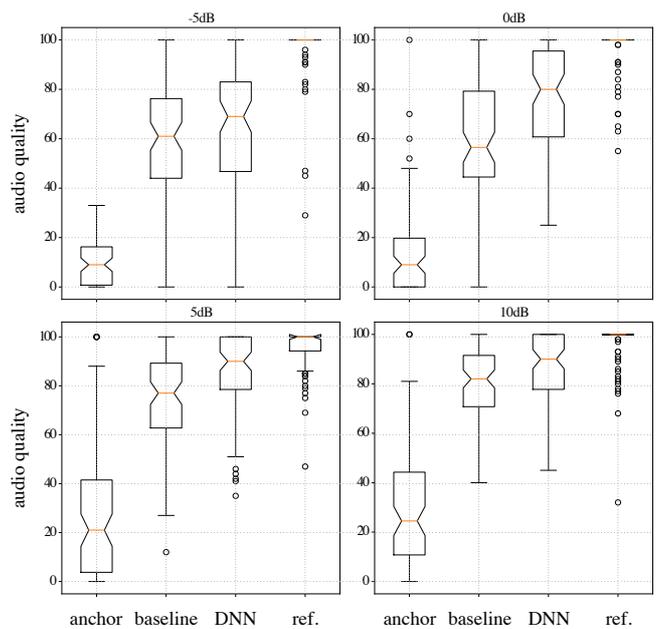}
	\caption{Subjective results (MUSHRA test, N=20) of the proposed algorithm ($\tau_1=200\,\mathrm{ms},\tau_2=2\,\mathrm{ms}$) against the minimum recursive tracking baseline and an ideal Wiener filter gain (ref). }
	\label{subjective}
\end{figure}

Looking at the results (Fig.\ref{subjective}), we find that the median subjective quality rating improves in all conditions over the baseline. Further, we find ceiling effects for the DNN-generated signal, indicating signals with equal quality to the reference for some conditions (cf. 5\,dB and 10\,dB).

\section{Summary}
In this work we presented a deep learning-based approach for noise reduction that is able to work with the restrictive conditions of hearing instrument signal processing. The subjective and objective evaluation showed improvements over a recursive minimum tracking approach in a wide range of SNR conditions using realistic noise scenarios.

\end{document}